\documentclass[aps,pra,amsmath,amssymb,twocolumn,superscriptaddress,showpacs,floatfix]{revtex4-1}

\usepackage{graphicx}
\usepackage{dcolumn}
\usepackage{bm}
\usepackage{subfigure}
\usepackage{float}
\usepackage{color}
\usepackage{amsmath}
\begin{document}

\title{Elementary Excitations in the Symmetric Spin--Orbital Model}

\author{M.~Y.~Kagan}
\affiliation{Kapitza Institute for Physical Problems, Russian Academy of Sciences, Moscow, 119334 Russia}
\affiliation{National Research University Higher School of Economics, Moscow, 109028 Russia}

\author{K.~I.~Kugel}
\affiliation{Institute for Theoretical and Applied Electrodynamics, Russian Academy of Sciences, Moscow, 125412 Russia}

\author{A.~V.~Mikheyenkov}
\affiliation{Institute for High Pressure Physics, Russian Academy of Science, Troitsk 142190, Russia}
\affiliation{Moscow Institute of Physics and Technology, Dolgoprudnyi, Moscow region, 141700 Russia}

\author{A.~F.~Barabanov}
\affiliation{Institute for High Pressure Physics, Russian Academy of Science, Troitsk 142190, Russia}
\date{\today }

\begin{abstract}
Possible types of elementary excitations in the symmetric spin--orbital model on the a square lattice are analyzed using a spherically symmetric self-consistent approach. The excitation spectra are calculated. The behavior of the corresponding correlation functions depending on the temperature and parameters of the model is studied. A schematic phase diagram is plotted. It is shown that the thermodynamics of the system is mainly determined by elementary excitations with the entangled spin and orbital degrees of freedom.
\end{abstract}

\maketitle

{\bf 1. Introduction.}
In the strongly correlated electron systems, such as transition metal compounds, a strong interplay of spin, orbital, and charge degrees of freedom leads to a rich variety of phase diagrams and to quite unusual phenomena, e.g., to colossal magnetoresistance~\cite{_Dagotto_Nanoscale_02,_Horsch_OrbPhys_07,
Khomsk05_PS,Kagan01_UFN}.

In the undoped compounds (magnetic insulators), the magnetism stems from the Anderson superexchange interaction~\cite{Anders59_PR} related to the virtual intersite hopping of electrons and leading to the Heisenberg-type Hamiltonian. In the systems with the orbital ordering, the magnetism is usually described in terms of the spin--orbital models. In such models, the Heisenberg-type spin interaction is supplemented by the interorbital coupling, as well as the coupling of the spin and orbital subsystems. Therefore, the orbitals determine the interaction between spins and @vice versa@. In real compounds, however, the exchange mechanism of the orbital interactions coexists with the purely lattice (Jahn--Teller) mechanism~\cite{_Kaplan_Cooperat_95}. At the fixed occupation of orbitals, the value magnitude and sign of the exchange interaction are determined by the Goodenough--Kanamori--Anderson rules~\cite{_GoodenoughMagn_63}. From the formal viewpoint, such spin--orbital physics could be interpreted as a generalization of the Heisenberg model, where the exchange integrals themselves are operators depending on the orbital degrees of freedom~\cite{KugelJETP_73,Kugel82_UFN}. As a result, the spin and orbital interactions can be of both ferro- and antiferromagnetic types. In such a situation, the exchange interaction turns out to be strongly frustrated even on square and cubic lattices. This leads to an enhancement of quantum effects in such systems~\cite{Oles00_PRB}. Thus, spins and orbitals form an entangled quantum state, which recently has awakened much interest (see, e.g., review article~\cite{Oles12_JoPCM}). These quantum effects are especially pronounced in the case of a square lattice~\cite{Wang09_PRB,Brzezi12_PRL}. The elementary excitations characteristic of orbital systems, namely, orbital waves or orbitons (similar to spin waves in magnetically ordered materials), also form a widely discussed topic~\cite{Brink98_PRB,Schlap12_N}. Nevertheless, in spite of the growing interest in orbital physics and a large number of publications in this field, the role of quantum entanglement for spin--orbital excitations and their contribution to the thermodynamics of the corresponding systems have not been properly addressed up to now. We consider this problem using as an example the symmetric spin--orbital model on a square lattice. We base our analysis on the spherically symmetric self-consistent approach providing reliable results for low-dimensional spin systems~\cite{Bara11_TMP_R,Mikh13_JL_R}. We demonstrate below that this approach is efficient for revealing the quantum entanglement of the spin and orbital degrees of freedom and the related correlation effects manifesting themselves even in the absence of a long-range order.

{\bf 2. Formulation of the problem.}
We proceed from the symmetric version of the spin--orbital model (sometimes referred to as the Kugel--Khomskii model) on a quadratic lattice~\cite{Kugel82_UFN}. The corresponding Hamiltonian has the form
\begin{equation*}
\widehat{\mathbf{H}}=\frac{J}{2}\sum_{\mathbf{i},\mathbf{g}}
{\mathbf{S}}_{\mathbf{i}}{\mathbf{S}}_{\mathbf{ig}}+\frac{I}{2}
\sum_{\mathbf{i},\mathbf{g}}{\mathbf{T}}_{\mathbf{i}}{\mathbf{T
}}_{\mathbf{ig}}+\frac{K}{2}\sum_{\mathbf{i},\mathbf{g}}
\left({\mathbf{S}}_{\mathbf{i}}{\mathbf{S}}_{\mathbf{ig}}\right)
\left( {\mathbf{T}}_{\mathbf{i}}{\mathbf{T }}_{\mathbf{ig}}\right),
\end{equation*}\begin{equation}
\widehat{\mathbf{H}}=\widehat{J}+\widehat{T}+\widehat{K} ,  \label{Hamilt}
\end{equation}
where $\mathbf{g}$ are the vectors for the nearest neighbor sites; $\mathbf{ig=i+g}$; $\widehat{\mathbf{S}}_{\mathbf{i}}$ and $\widehat{\mathbf{T}}_{\mathbf{i}}$ are the spin and pseudospin operators (the latter describes the orbital degrees of freedom), respectively; $S=1/2$ and $T=1/2$.

We consider the case of the antiferromagnetic (AFM) interaction within each subsystem, $J=I>0$, and of the negative intersubsystem exchange, $K<0$ (further on, all energies are given in the units of $J=I=1$). The earlier analysis of the one-dimensional symmetric spin--orbital model~\cite{Pati98_PRL} suggests that, just at such relations between the parameters, the effects of the coupling between the spin and orbital degrees of freedom are most clearly pronounced. Moreover, the entanglement of the spin and orbital excitations also manifests itself quite clearly~\cite{You12_PRB,Lundgr12_PRB}.

Note also that a generally recognized description of an antiferromagnet even for one subsystem ($K=0$) does not exist in the two-dimensional (2D) case at nonzero temperatures. Below, we use one of the versions of the mean-field approximation, namely, the spherically symmetric self-consistent approach (SSSA) (see, e.g., \cite{Bara11_TMP_R,Kondo72_PTP,Shimah91_JPSJ}. The characteristic feature of such an approach is the possibility of finding the temperature dependence of the correlation functions $\langle\widehat{{S}}_{\mathbf{i}}\widehat{{T}}_{\mathbf{i}}\rangle$ describing the coupling between subsystems. In this approach, the condition of the spherical symmetry (and hence the Mermin--Wagner theorem~\cite{Mermin66_PRL}) is explicitly met, all sites in the system are equivalent, one-site averages vanish
\begin{equation}
\langle\widehat{{S}}_{\mathbf{i}}\rangle =\langle \widehat{{T
}}_{\mathbf{i}}\rangle =0 \label{site_aver}
\end{equation}
and the correlation functions for different components of spin and pseudospin ($\alpha \neq \beta$) are also equal to zero
\begin{equation}
\langle \widehat{S}_{\mathbf{i}}^{\alpha}\widehat{S}_{\mathbf{j}}^{\beta}\rangle=0, \quad
\langle \widehat{T}_{\mathbf{i}}^{\alpha}\widehat{T}_{\mathbf{j}}^{\beta}\rangle=0, \quad
\langle \widehat{S}_{\mathbf{i}}^{\alpha}\widehat{T}_{\mathbf{j}}^{\beta}\rangle=0 . \label{ini_corrs}
\end{equation}

{\bf 3. $K=0$ case.}
In the absence of coupling between subsystems, i.e., at $K=0$, in the framework of SSSA, the spin--spin Green's function $G^{zz}\left(\omega,
\mathbf{q}\right)=\langle S_{\mathbf{q}}^{z}|S_{-\mathbf{q}}^{z}\rangle _{\omega },\  (G^{zz}=G^{xx}=G^{yy})$ is known to have the form \cite{Bara11_TMP_R,Shimah91_JPSJ}
\begin{equation}
G^{zz}\left( \omega ,\mathbf{q}\right) =\frac{F_{\mathbf{q}}}{\omega
^{2}-\omega _{\mathbf{q}}^{2}}  \label{Gzz0}\, ,
\end{equation}
where the numerator is determined by the expressions
\begin{equation}
F_{\mathbf{q}}=-8J(1-\gamma _{\mathbf{q}})c_{g} \label{Fq0},
\end{equation} \begin{equation}
\gamma_{\mathbf{q}}=\frac{1}{4}\sum_{\mathbf{g}}e^{i\mathbf{qg}}=
\frac{1}{2}(\cos (q_{x})+\cos (q_{y})) \label{gamq0},
\end{equation}
$c_{g}=\langle \widehat{S}_{\mathbf{i}}^{z}\widehat{S}_{\mathbf{i+g}}^{z}\rangle$
is the spin--spin correlation function for the nearest-neighbor sites.

The spin excitation spectrum $\omega_{\mathbf{q}}$ can be written in the form
\begin{equation}
\omega _{\mathbf{q}}^{2}=2J^{2}(1-\gamma_{\mathbf{q}})
\Bigl\{1+4\bigl[\widetilde{c}_{2g}+2\widetilde{c}_{d}-
\widetilde{c}_{g}(1+4\gamma _{\mathbf{q}})\bigr]\Bigr\}. \label{wq20}
\end{equation}
In the approximation of one vertex correction $\alpha$, the correlation functions in \eqref{wq20} are $\widetilde{c}_{r}=\alpha c_{r}$ \cite{Bara11_TMP_R,Mikh13_JL_R,Hartel11_PRB}.
Three correlation functions $c_{r}$ (where $r=g, d$ and $2g$ correspond to the first, second, and third nearest neighbors, respectively) and the vertex corrections are determined self-consistently in terms of the Green's function $G^{zz}$ under the additional constraint $\langle \widehat{\mathbf{S}}_{\mathbf{i}}^{2}\rangle =3/4$.

In contrast to the approaches assuming the two-sublattice ground state, the Brillouin zone points $\mathbf{\Gamma}=(0,0)$ and $\mathbf{Q}=(\pi,\pi)$ in the SSSA are nonequivalent. At $T\neq0$ (and $K=0$), there is no AFM long-range order in either of the subsystems. In the spin excitation spectrum \eqref{wq20}, a gap at the $\mathbf{Q}$ point is open, $\omega _{\mathbf{Q}}>0$, and the spin--spin correlation function vanishes at infinity. At $T\rightarrow 0$, the gap at $\mathbf{Q}$ closes and the spin--spin correlation function has a nonzero value at infinity (it changes sign according to the "Manhattan length"\ rule, i.e., at the displacements by the lattice constant along the horizontal or vertical direction). This implies the formation of a long-range AFM order. The same is also valid for the pseudospin subsystem.

{\bf 4. Nonzero $S-T$ interaction.}
Let us consider the structure of the third term in \eqref{Hamilt}
\begin{eqnarray}
\widehat{K}&=&\frac{K}{2}\sum_{\mathbf{i,g}}S_{\mathbf{i}}^{\alpha }S_{
\mathbf{ig}}^{\alpha }T_{\mathbf{i}}^{\beta }T_{\mathbf{ig}}^{\beta
}=\sum_{\langle \mathbf{i,j\rangle }}\widehat{K}_{\mathbf{i,j}} , \\
\widehat{K}_{\mathbf{i,j}}&=&KS_{\mathbf{i}}^{\alpha }S_{\mathbf{j}}^{\alpha
}T_{\mathbf{i}}^{\beta }T_{\mathbf{j}}^{\beta };\ (\alpha ,\beta =x,y,z) ,
\label{Knm1}
\end{eqnarray}
where $\langle \mathbf{i,j} \rangle$ means the summation over the nearest-neighbor bonds.

Taking into account the spherical symmetry (conditions \eqref{site_aver} and \eqref{ini_corrs}), the Hamiltonian $\widehat{K}$ in the mean-field representation allows for separating out nonzero averages of the $\langle \widehat{S}\widehat{S}\rangle$, $\langle \widehat{T}\widehat{T}\rangle$, and $\langle \widehat{S}\widehat{T}\rangle$ types. The separating out of averages of the $\langle \widehat{S}\widehat{S}\rangle$ and $\langle \widehat{T}\widehat{T}\rangle$ types leads only to a renormalization of parameters $I$ and $J$, which do not involve mixing of the subsystems. We omit them below. Then, $\widehat{K}$ has the mean-field representation
\begin{equation}
\widehat{K}_{\mathbf{i,j}}\approx a\,\hat{K}_{\mathbf{i,j}}^{a}+b\hat{K}_{
\mathbf{i,j}}^{b} , \label{Kij0}
\end{equation}
\begin{eqnarray}
\hat{K}_{\mathbf{i,j}}^{a} \!\!&=&\!\!S_{\mathbf{i}}^{\alpha}{T}_{\mathbf{i}}^{\alpha}\langle
S_{\mathbf{j}}^{\alpha}{T}_{\mathbf{j}}^{\alpha}\rangle \!\!+\!\! \langle S_{\mathbf{i}}^{\alpha}{T}
_{\mathbf{i}}^{\alpha}\rangle S_{\mathbf{j}}^{\alpha}{T}_{\mathbf{j}}^{\alpha}\!\!-\!\!\langle S_{
\mathbf{i}}^{\alpha}{T}_{\mathbf{i}}^{\alpha}\rangle\! \langle S_{\mathbf{j}}^{\alpha}{T}_{
\mathbf{j}}^{\alpha}\rangle \label{Kij_a} \\
\hat{K}_{\mathbf{i,j}}^{b} \!\!&=&\!\!S_{\mathbf{i}}^{\alpha}{T}_{\mathbf{j}}^{\alpha}\langle
S_{\mathbf{j}}^{\alpha}{T}_{\mathbf{i}}^{\alpha}\rangle \!\!+\!\! \langle S_{\mathbf{i}}^{\alpha}{T}
_{\mathbf{j}}^{\alpha}\rangle S_{\mathbf{j}}^{\alpha}{T}_{\mathbf{i}}^{\alpha}\!\!-\!\!\langle S_{
\mathbf{i}}^{\alpha}{T}_{\mathbf{j}}^{\alpha}\rangle\! \langle S_{\mathbf{j}}^{\alpha}{T}_{
\mathbf{i}}^{\alpha}\rangle   \label{Kij_b}
\end{eqnarray}
At the same time, the condition $a+b=1$ should be met because in the mean-field representations we on the same footing neglect the terms $\langle \delta S\,\delta {S}\rangle $, $\langle \delta T\,\delta {T}\rangle $, and $\langle \delta S\,\delta {T}\rangle $, which are quadratic in terms of the fluctuations. In Eqs.\eqref{Kij_a} and \eqref{Kij_b}, the terms off-diagonal with respect to the Greek superscripts drop out according to conditions \eqref{ini_corrs}.

Case \eqref{Kij_a} corresponds to separating out the on-site averages, whereas case \eqref{Kij_b} describes separating out the averages related to the operators corresponding to the neighboring sites. From the viewpoint of the underlying physics, it is clear that version $a$ specified by Eq. \eqref{Kij_a} gives the dominant contribution to the final results. It corresponds to the on-site $\mathbf{S}$ and $\mathbf{T}$ correlations, which are determined by the on-site Coulomb interactions and appear to be more significant.

Note that, in the formal limit of $I=J=0$, $K\neq 0$, these cases correspond to quite different states of the system. At $a=1$ and $b=0$, the system breaks up into the on-site noninteracting "dimers"\ composed of $\hat{\mathbf{S}}$ and $\hat{\mathbf{T}}$ operators. Switching on $I=J>0$ loosens the dimers.

In the case of $a=0$ and $b=1$, the system breaks up into two "checkerboard"\ subsystems; in each of them, one sublattice contains the $\hat{\mathbf{S}}$ operators and the other is occupied by the $\hat{\mathbf{T}}$ operators. In such situation (in the limit of $I=J=0$, $K\neq 0$), the ferromagnetic order arises in each subsystem. At $I=J>0$, the coupling between the subsystems is switched on.

All further calculations are performed taking into account both types of contributions, \eqref{Kij_a} and \eqref{Kij_b}. Then, the energy is minimized with respect to both coefficients, $a$ and $b$. Within the whole parameter range under study, this results in the validity of the $a=1$ and $b=0$ limit. For the sake of brevity, all expressions are represented below just in this limit, i.e., taking into account the terms in form \eqref{Kij_a}.

\begin{figure}[tbp]
\begin{center}
\includegraphics[width=\columnwidth]{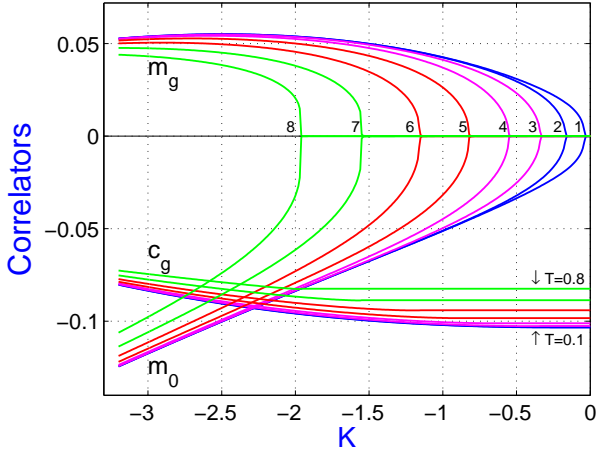}
\caption{Spin--spin correlation function $c_{g}$ for the nearest-neighbor sites and also spin--pseudospin correlation functions $m_{0}$ and $m_{g}$ for the first and second coordination spheres versus the temperature and intersubsystem exchange parameter $K$.
The curves forming a "platypus nose"\ are $m_{0}(K)$ ($m_{0}<0$) and $m_{g}(K)$ ($m_{g}>0$). The numbers from $1$ to $8$ enumerate the temperature $T=0.1\div0.8$, respectively. Lower curves depict the behavior of $c_{g}$ (we indicate the boundary values of $T$).
}
\end{center}
\end{figure}
\begin{figure}[tbp]
\begin{center}
\includegraphics[width=\columnwidth]{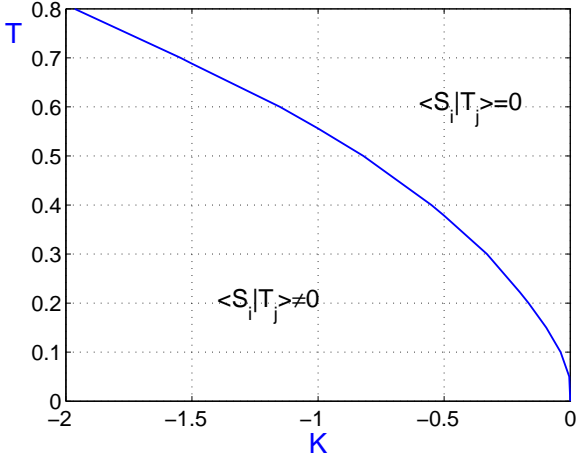}
\caption{Regions corresponding to zero and nonzero spin--pseudospin correlations. The phase boundary is well fitted by the $T_c=0.55|K|^{0.55}$ curve.}
\end{center}
\end{figure}

{\bf 5. Correlation functions and spectrum at $K\neq0$.}
At nonzero coupling between the subsystems, it is necessary to introduce both spin--spin and spin--pseudospin Green's functions
\begin{equation}
G_{q}^{zz}=\left\langle S_{\mathbf{q}}^{z}\mid
S_{\mathbf{-q}}^{z}\right\rangle _{\omega } , \label{Gq0}
\end{equation}
\begin{equation}
R_{\mathbf{q}}^{zz}=\left\langle T_{\mathbf{q}}^{z}\mid
S_{\mathbf{-q}}^{z}\right\rangle _{\omega } , \label{Rq0}
\end{equation}
Here, $\left\langle T_{\mathbf{q}}^{z}\mid T_{\mathbf{-q}}^{z}\right\rangle _{\omega }
=\left\langle S_{\mathbf{q}}^{z}\mid S_{\mathbf{-q}}^{z}\right\rangle _{\omega }$, since we consider the symmetric case $I=J$.

The calculations within the conventional SSSA scheme lead to the following expressions for $G_{q}^{zz}$ and $R_{\mathbf{q}}^{zz}$:
\begin{equation}
G_{q}^{zz}=\frac{F_{ac}(\mathbf{q})}{\omega ^{2}-\omega
_{ac}^{2}(\mathbf{q})}+\frac{F_{opt}(\mathbf{q})}{\omega ^{2}-\omega
_{opt}^{2}(\mathbf{q})} , \label{Gq}
\end{equation}
\begin{equation}
R_{\mathbf{q}}^{zz}=\frac{F_{ac}(\mathbf{q})}{\omega ^{2}-\omega
_{ac}^{2}(\mathbf{q})}-\frac{F_{opt}(\mathbf{q})}{\omega ^{2}-\omega
_{opt}^{2}(\mathbf{q})} , \label{Rq}
\end{equation}
where
\begin{equation}
F_{ac}=\frac{F_{1}+F_{2}}{2},\ F_{opt}=\frac{F_{1}-F_{2}}{2},
\end{equation} \begin{equation}
F_{1}=-8Jc_{g}(1-\gamma _{\mathbf{q}})-Mm_{0}, \ F_{2}=Mm_{0}, \label{F12}
\end{equation}
The excitation spectra have the form
\begin{equation}
\omega_{ac}^{2}(\mathbf{q})=W+Z,\ \omega _{opt}^{2}(\mathbf{q})=W-Z, \label{spectra}
\end{equation} \begin{eqnarray}
W&=&2J^{2}(1\!-\!\gamma _{\mathbf{q}})\Bigl\{1
+4\bigl[\widetilde{c}_{2g}\!+\!2\widetilde{c}_{d}\!-\!\widetilde{c}_{g}(1\!+\!4\gamma _{\mathbf{q}})\bigr]\Bigr\}\!+\notag \\
&&+4JM(2\widetilde{m}_{g}-\widetilde{m}_{g}\gamma_{\mathbf{q}}-\widetilde{m}_{0}\gamma _{\mathbf{q}})+\frac{1}{8}M^{2}, \label{W} \\
Z&=&-4JM\left[ \widetilde{c}_{g}(1-\gamma _{\mathbf{q}})+\widetilde{m}_{g}-\widetilde{m}_{0} \gamma_{\mathbf{q}}\right]
 -\frac{1}{8}M^{2}, \label{Z}
\end{eqnarray}

\begin{figure}[tbp]
\begin{center}
\includegraphics[width=\columnwidth]{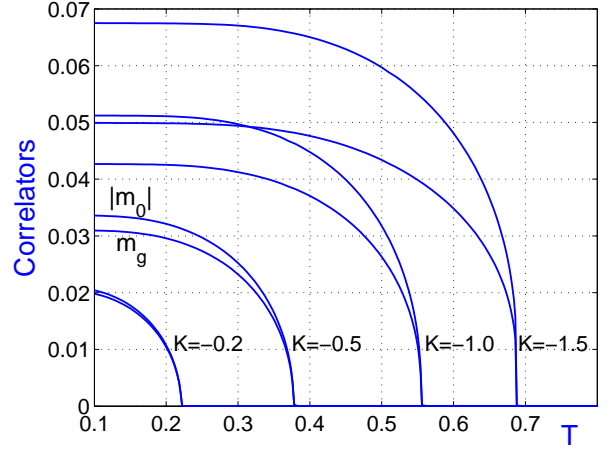}
\caption{Temperature dependence of the spin--pseudospin correlation functions $m_{0}$ and $m_{g}$ at several fixed $K$ values. In each pair of curves, the upper and lower curves correspond to $|m_{0}|$ and $m_{g}$, respectively. All curves are well fitted by the power law $m\sim (T_c-T)^\alpha$ with the exponent $\alpha \sim 0.3\div0.5$ nearly independent of $K$.}
\end{center}
\end{figure}
Here, we have $M=8Km_{0}$, $\gamma _{\mathbf{q}}$ is defined in \eqref{gamq0}, and $\widetilde{c}_{g}$, $\widetilde{c}_{d}$, and $\widetilde{c}_{2g}$ are again the spin--spin correlation functions for the first three coordination spheres (taking into account the vertex corrections). Expressions \eqref{W} and \eqref{Z} also include the on-site ($m_{0}$) and intersite ($m_{g}$) spin--pseudospin correlation functions
\begin{equation}
m_{0}=\langle S_{\mathbf{i}}^{z}T_{\mathbf{i}}^{z}\rangle,\ \
m_{g}=\langle S_{\mathbf{i}}^{z}T_{\mathbf{ig}}^{z}\rangle ,
\end{equation}
and for the corresponding vertex corrections in
$\widetilde{m}_{0}=\alpha _{ST}^{0}m_{0}$ and 
$\widetilde{m}_{g}=\alpha _{ST}^{g}m_{g}$
we use the simplest approximation
$\alpha _{ST}^{0}=\alpha _{ST}^{g}=1$.

The following relationships are always valid:
\begin{equation}
\omega_{opt}(\mathbf{\Gamma})\geq\omega_{ac}(\mathbf{\Gamma})=0, \quad
\omega_{ac}(\mathbf{Q})\geq\omega_{opt}(\mathbf{Q})\geq 0. \label{omegas}
\end{equation}

The numerical procedure for $K\neq0$ is similar to that described above for $K=0$. All correlation functions ($c_{r}$ ($(r=g, d, 2g)$), $m_{0}$, and $m_{g}$) and vertex corrections involved in the problem are determined in a self-consistent manner using the Green's functions $G^{zz}$ and $R^{zz}$. The results presented below correspond to the case of nonzero temperatures.

{\bf 6. Results and discussion.}
In Fig.~1, we demonstrate the behavior of the spin--spin correlation function $c_{g}$ for the nearest-neighbor sites, as well as of the spin--pseudospin (spin--orbital) correlation functions $m_{0}$ and $m_{g}$ for the first and second coordination spheres depending on the temperature and intersubsystem exchange parameter $K$. We can see that, at a certain (temperature dependent) value of $K$, both on-site ($m_{0}$) and intersite ($m_{g}$) spin--orbital correlation functions become nonzero and their absolute values start growing steeply. At the same time, the spin--spin correlation function $c_{g}$ varies smoothly without any peculiar features. This suggests the formation of an entangled state in the system, which is characterized by nonzero values of the spin--orbital correlation functions. The transition to such a state mimics a second-order phase transition. Nevertheless, here, strictly speaking, both the spin and orbital long-range orders are absent.

In Fig. 2, we show the phase diagram (the regions corresponding to zero and nonzero spin--pseudospin correlations). The boundary between these states is well fitted by a power-law dependence with the exponent close to $1/2$.

In Fig. 3, we illustrate the temperature dependence of the spin--pseudospin correlation functions $m_{0}$ and $m_{g}$ at several fixed $K$ values. These curves are also well fitted by the power law $m\sim (T_c-T)^\alpha$ with the exponent $\alpha \sim 0.3\div0.5$ nearly independent of $K$. Nonzero values of $m_{0}$ and $m_{g}$ arising at $T\gtrsim T_c$ (nearly indistinguishable at the scale of the figure) characterize the accuracy of the self-consistent calculations.

In Figs. 4 and 5, we demonstrate the elementary excitation spectra, $\omega_{ac}(\mathbf{q})$ and $\omega_{opt}(\mathbf{q})$ \eqref{spectra}, at temperature $T=0.3$ for the cases of the weak ($K=-0.4$) and strong ($K=-3.0$) coupling between the subsystems. In the former case, the splitting in the spectrum is rather small and can be seen only in the vicinity of the high-symmetry points $\mathbf{\Gamma}=(0,0)$ and $\mathbf{Q}=(\pi,\pi)$. In the latter case, it is so clearly pronounced that the upper parts of the branches form a nearly dispersionless region. The spectrum is split at $T<T_c$, and this is true for both cases (as follows from the plot in Fig.~2). In Figs.~4 and 5, we can see that relationships \eqref{omegas} are met. In particular, the $\omega_{ac}(\mathbf{\Gamma})$ branch always corresponds to the Goldstone mode. The splitting between the branches grows with $|K|$ and decreases with the growth of $T$.

\begin{figure}[tbp]
\begin{center}
\includegraphics[width=\columnwidth]{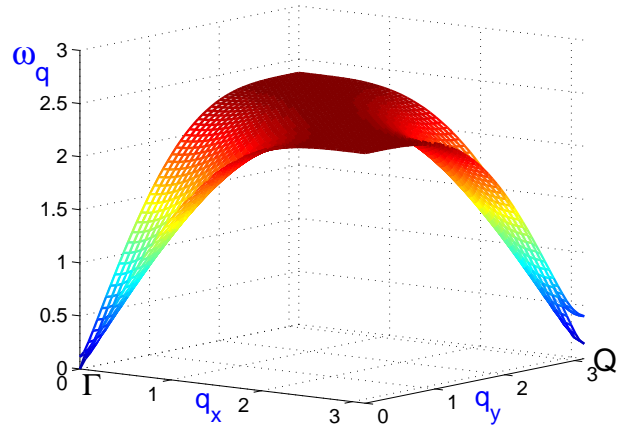}
\caption{Elementary excitation spectra, $\omega_{ac}(\mathbf{q})$ and $\omega_{opt}(\mathbf{q})$ \eqref{spectra}, at temperature $T=0.3$ for a weak splitting ($K=-0.4$). Only the first quarter of the Brillouin zone is shown.}
\end{center}
\end{figure}
\begin{figure}[tbp]
\begin{center}
\includegraphics[width=\columnwidth]{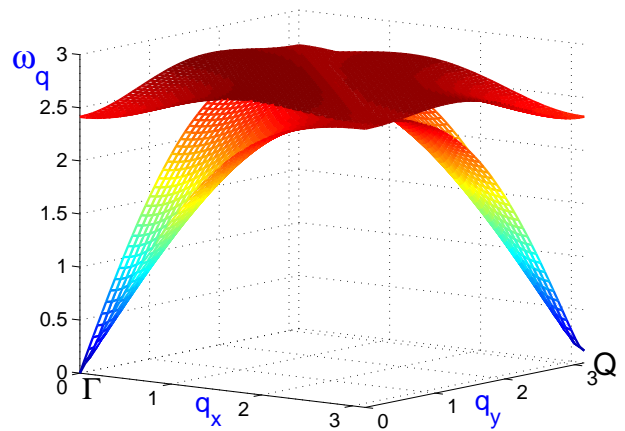}
\caption{Elementary excitation spectra, $\omega_{ac}(\mathbf{q})$ and $\omega_{opt}(\mathbf{q})$ \eqref{spectra}, at temperature $T=0.3$ for a strong splitting ($K=-3.0$). Only the first quarter of the Brillouin zone is shown. The upper parts of the spectral branches form a nearly dispersionless region. The scale on the coordinate axes is the same as in Fig. 4.
}
\end{center}
\end{figure}
At $K=0$, the $\mathbf{S}$ and $\mathbf{T}$ subsystems are independent and (since we are dealing with the case of $I=J$) the excitation spectra are the same in both subsystems, $m_{0}=m_{g}=0$. At a sufficiently large $K \neq 0$, the degenerate excitation spectrum splits into two branches. This means that, in the measured magnetic susceptibility $\chi(q,\omega)$, one should observe an additional peak related to the interaction between the two subsystems.

{\bf 7. Conclusions.}
Thus, we have demonstrated that the symmetric spin--orbital model exhibits the formation of the state with nonzero values of the correlation functions corresponding to the entanglement of the spin and orbital degrees of freedom. The transition to such a state resembles by its characteristics a second-order phase transition.
It would be interesting to analyze the characteristic features of this entangled state also for more realistic models (taking into account the anisotropy, other relations between the exchange integrals, etc.).
This work was supported by the Russian Foundation for Basic Research (project nos. 13-02-00909-a, 14-02-00058, and 14-02-00278).

\end{document}